\documentclass[12pt,preprint]{aastex}

\shorttitle{A New Model For The Loop-I (The North Polar Spur) Region}
\shortauthors{M. Wolleben}

\begin{document}

\title{A New Model For The Loop-I (The North Polar Spur) Region}

\author{M. Wolleben\altaffilmark{1,2}}

\altaffiltext{1}{Department of Electrical and Computer Engineering, University of Alberta, Edmonton, Alberta, Canada T6G 2V4.}

\altaffiltext{2}{National Research Council Canada, Herzberg Institute of Astrophysics, Dominion Radio Astrophysical Observatory, P.O. Box 248, Penticton, BC, V2A 6J9, Canada; maik.wolleben@nrc-cnrc.gc.ca}

\begin{abstract}

The North Polar Spur (NPS) is the brightest filament of Loop I, a large circular feature in the radio continuum sky.  In this paper, a model consisting of two synchrotron emitting shells is presented that reproduces large-scale structures revealed by recent polarization surveys. The polarized emission of the NPS is reproduced by one of these shells. The other shell, which passes close to the Sun, gives rise to polarized emission towards the Galactic poles. It is proposed that X-ray emission seen towards the NPS is produced by interaction of the two shells. Two OB-associations coincide with the centers of the shells. A formation scenario of the Loop~I region is suggested. 

\end{abstract}


\section{Introduction}

Total intensity surveys reveal a number of radio spurs that can be joined into small circles on the sky, so-called radio loops. One of these is Loop~I, which has an intriguing filament called the North-Polar Spur (NPS). It has been concluded by several authors \citep[e.g.][and references therein]{1971A&A....14..252B, 1979ApJ...229..533H, 1983BASI...11....1S} that the radio loops are correlated with expanding gas and dust shells, energized by supernovae or stellar winds. Accordingly, the Loop~I superbubble was attributed to stellar winds from the SCO-CEN OB association and supernova activity in the same vicinity, with the NPS being the brightest segment of a supernova remnant (SNR). Its magnetic field (B-field) has been modelled by \citet{1972A&A....21...61S} and \citet{1998LNP...506..229H} based on radio and optical polarization data, suggesting that the local B-field is deformed by an expanding shell. According to a model by \citet{1979IAUS...84..295W}, an SNR, produced by a member of the SCO-CEN association, has expanded inside the Loop~I bubble. Its shell is just beginning to encounter and interact with the surface of the surrounding \ion{H}{1} shell (GS 331+14-15). There has been a debate over the expansion velocity ($v_{\mathrm{exp}}$) of GS 331+14-15: Weaver gives $v_{\mathrm{exp}}\approx 2$~km$\,$s$^{-1}$, based on data from the northern hemisphere, while \citet{1984ApJS...55..585H} and \citet{1974PASJ...26..399S} find a higher $v_{\mathrm{exp}}$ of $19$ and $25$~km$\,$s$^{-1}$. In a picture by \citet{1987Ap&SS.138..229B} the \ion{H}{1} shell has already passed the Sun.

The relatively low expansion velocity of the neutral gas surrounding Loop~I implies an age of some $10^6$ years and hence suggests the NPS to be part of an old SNR. At this age, however, an SNR is well beyond its radiative phase and its optical and radio emission become invisible. On the other hand, X-ray and radio emission have been observed towards the NPS, indicating an age at least one order of magnitude less.  Based on X-ray data, \citet{1977ApJ...218..511B} suggest that Loop~I is an old but recently reheated SNR. This was refined by \citet{1995ASPC...80...45E} who argue that Loop~I is caused by a shock from a recent supernova, heating the inner wall of the SCO-CEN supershell, and giving rise to the prominent X-ray feature of the NPS. 

Recently, \citet{2006A&A...448..411W} published the new DRAO Low-Resolution Polarization Survey of the northern sky at $1.4$~GHz\footnote{See also \citet[][]{2005PhDT.........3W}. The data are publicly available and can be downloaded at: \url{http://www.drao.nrc.ca/26msurvey}, \url{http://www.mpifr-bonn.mpg.de/div/konti/26msurvey}, or from CDS, Strasbourg \citep{2005yCat..34480411W}}, which is part of the Canadian Galactic Plane Survey \citep[][]{2003AJ....125.3145T}. Making use of these new data as well as the WMAP polarization data at 23~GHz (Page et al. 2006), a new and detailed study of the Loop~I region can be performed. In this paper I describe a model consisting of two interweaving magnetic shells, which explains some newly detected polarization and depolarization structures, correctly predicts the shape of the NPS, as well as providing a clue to the origin of X-ray emission observed towards the NPS. 

\section{Large Polarized Structures In The DRAO Survey} 

\begin{figure}
\includegraphics[width=15.5cm]{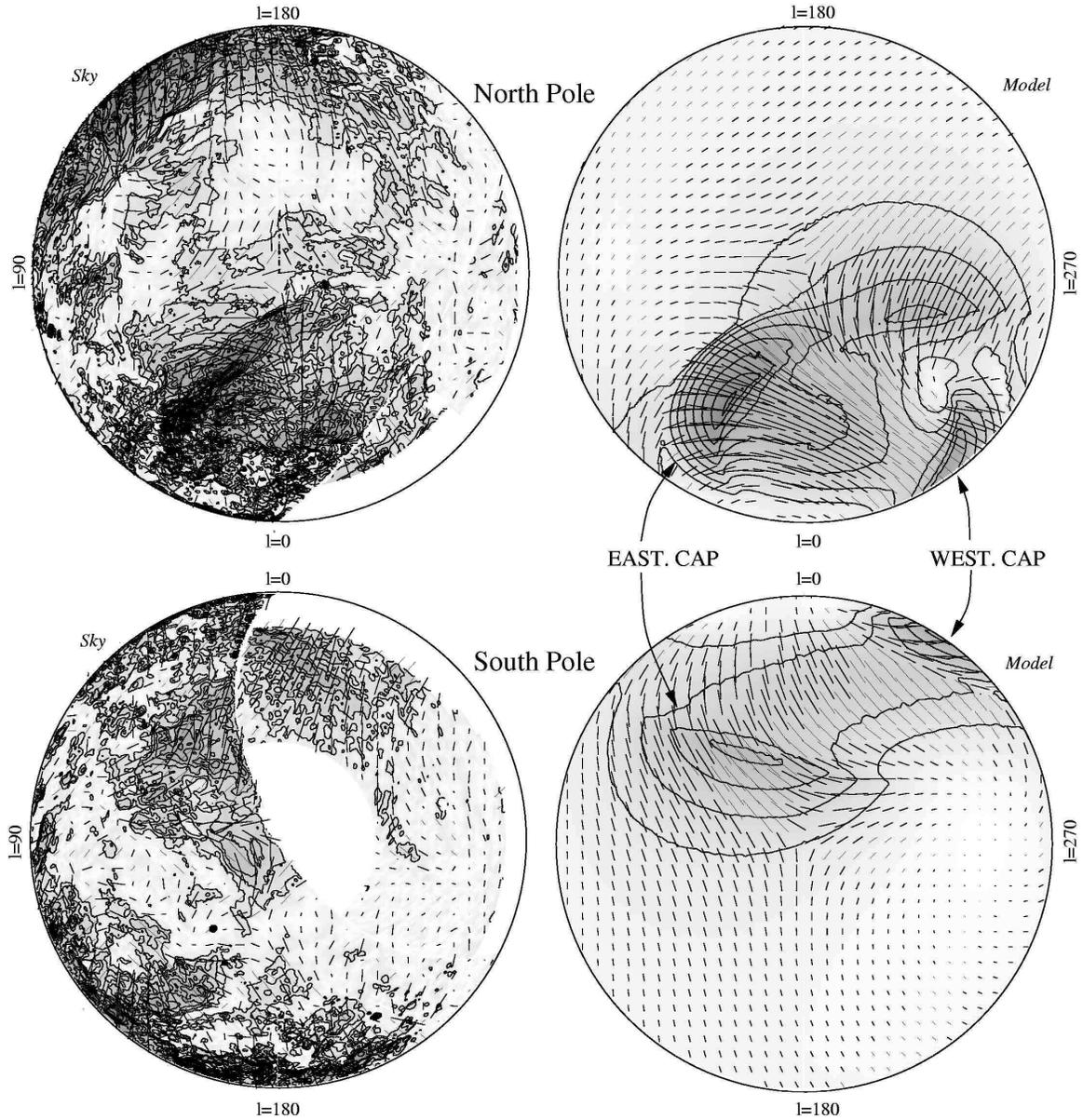}
\caption{Observed ({\it left}) and modelled ({\it right}) maps of the northern ({\it top}) and southern ({\it bottom}) Galactic poles with polarization vectors overlayed. The contours show polarized intensity from the DRAO and WMAP data from $50$ to $400$~mK, in steps of $50$~mk.}
\label{stereo}
\end{figure}

The polarization data used in this paper are shown in Figures~\ref{sky} and \ref{stereo} The NPS has long been known to be one of the most intensely polarized features of the sky. Its filaments cover large parts of the northern Galactic hemisphere at latitudes of $b\gtrsim 30\degr$, with an abrupt drop of polarized emission below $30\degr$ latitude at 1.4~GHz. Surprisingly, structures of the NPS are very evident in polarized intensity but its filamentary structure is not detectable in polarization angle. In the following, previously unknown polarization features detected in the DRAO polarization survey, which are of relevance for this study, are described:

\subsection{High Latitude Polarized Emission (HLPE)}

\begin{figure}
\includegraphics[width=15.5cm]{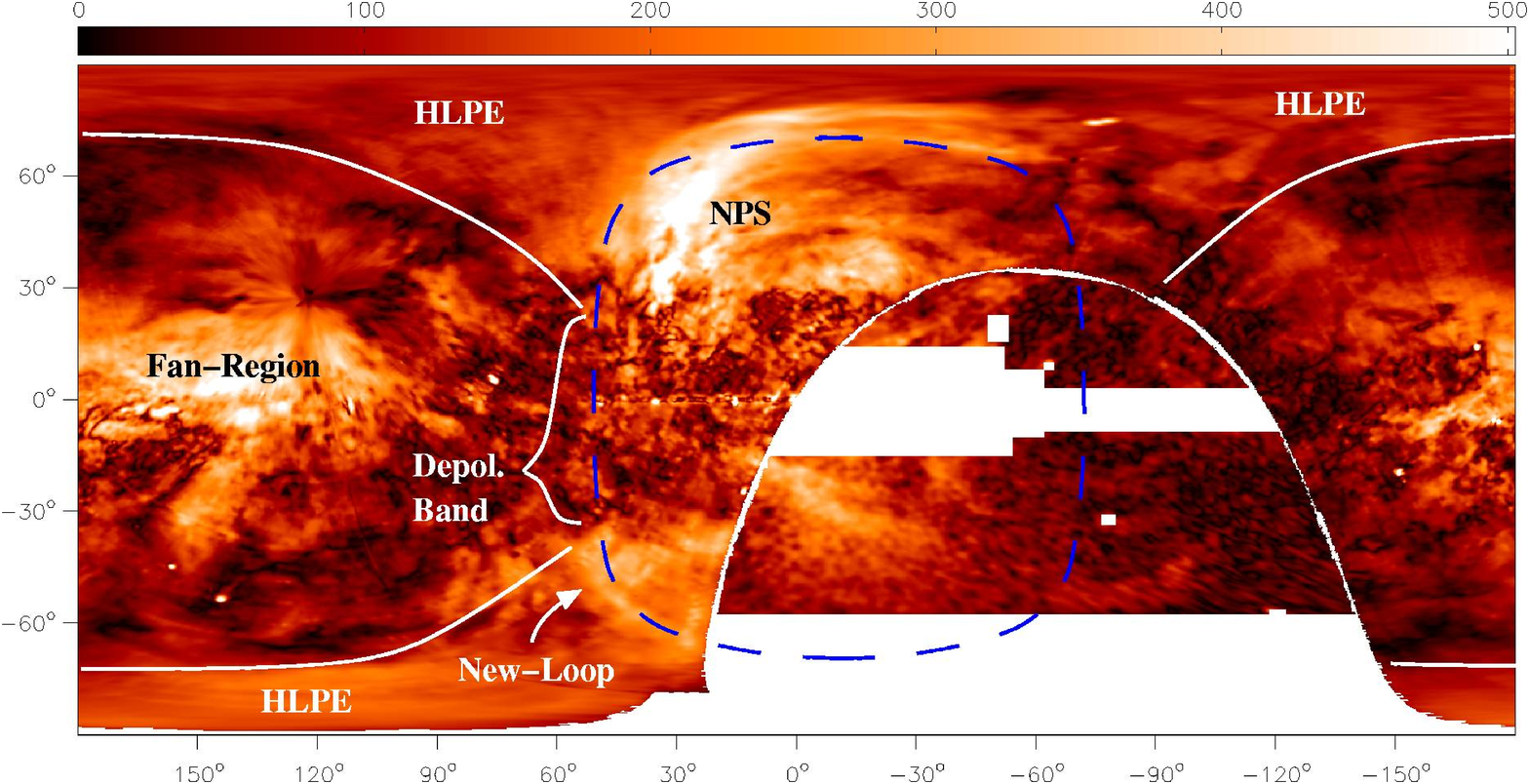}\\
\includegraphics[width=15.5cm]{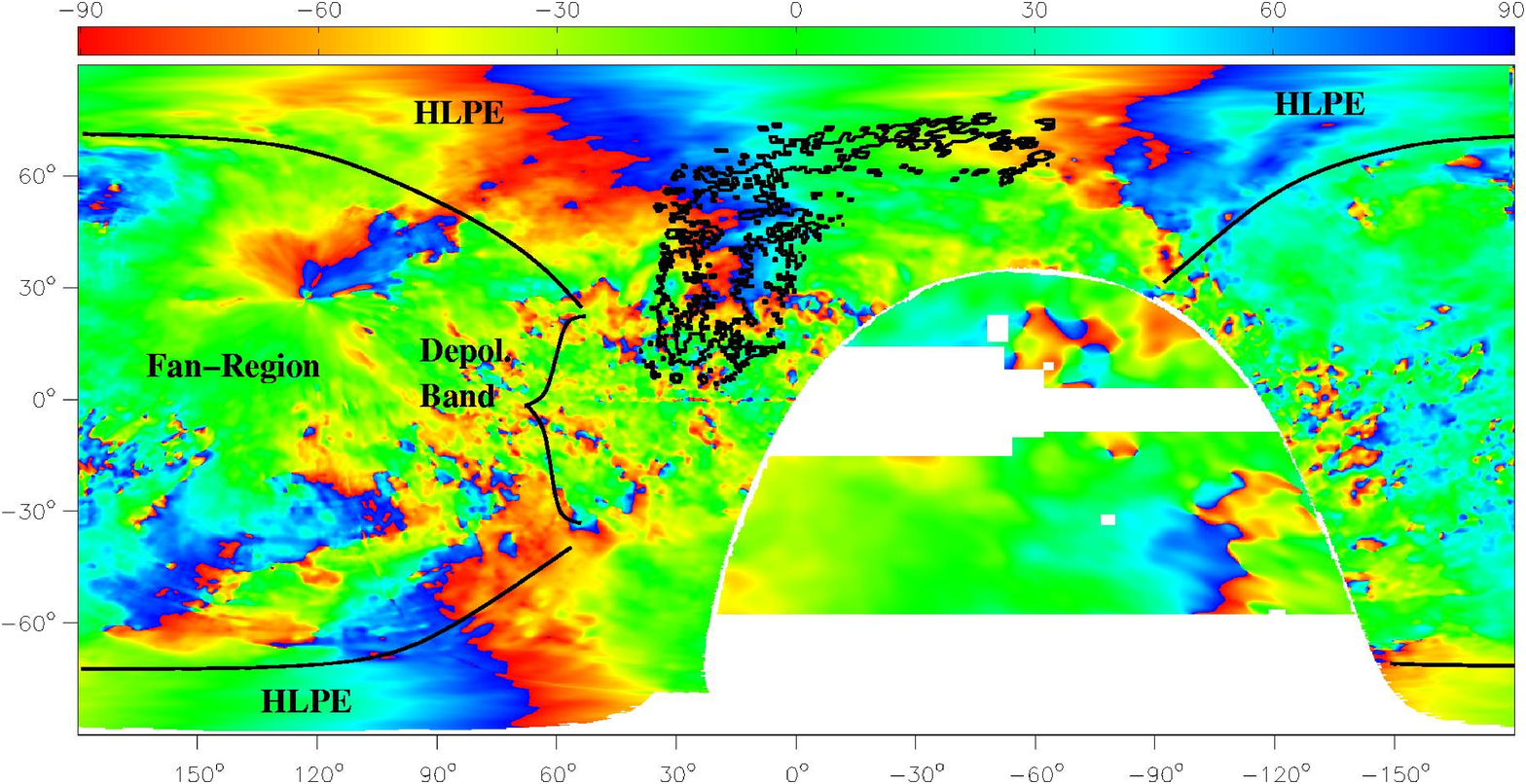}
\caption{Map of polarized intensity in units of mK ({\it top}) and polarization angle ({\it bottom}) in Galactic coordinates, taken from the DRAO polarization survey at $1.4$~GHz (northern hemisphere) and the WMAP polarization data smoothed to $4\degr$ resolution (southern hemisphere). The data are shown in rectangular projection to expose structures around the Galactic poles. The polarized intensity map exhibits three extended features: the NPS (from $l=310\degr$ to $50\degr$ and $b=30\degr$ to $80\degr$), the New Radio Loop (centered at $l=40\degr$, $b=-55\degr$, about $40\degr$ in diameter), and the so-called ``Fan-Region'' (centered at $l = 140\degr$, $b = 5\degr$, about $60\degr$ in diameter). The solid lines mark the observable extent of the HLPE in the second and third Galactic quadrant. The dashed line shows a circle fitted through the New Loop and one of the two extended ``arms'' of the NPS (at about $b=70\degr$). Black contours in the bottom panel show $1.5$~keV X-ray emission associated with the NPS at levels of $200$ and $350\times 10^{-6}$counts/s  \citep[taken from][]{1995ApJ...454..643S}.}
\label{sky}
\end{figure}

In Fig.~\ref{sky} (top), towards the Galactic poles ($b=\pm 90\degr$), a systematic pattern of polarized emission and polarization angle is visible (HLPE, hereafter).  
\label{tpexcess}
At $408$~MHz, the all-sky map of \citet{1982A&AS...47....1H} shows a weak counterpart of the HLPE in total intensity (Stokes-I) with excess emission at the Galactic poles of $\sim 4$~K (northern pole) and $\sim 5$~K (southern pole). The Stockert survey at $1420$~MHz \citep{1986A&AS...63..205R} also shows weak Stokes-I emission at the poles of $\sim 60$~mK (northern pole) and $\sim 200$~mK (southern pole). These temperatures imply a spectral index of $-3.4$ and $-2.6$ for the northern and southern HLPE, respectively, and thereby suggest that the HLPE is synchrotron emission. 

What could cause enhanced synchrotron emission around the Galactic poles -- the HLPE? 

First, the HLPE could be (polarized) synchrotron emission from the local spiral arm, which would produce a similar pattern of polarization angles towards the Galactic poles. Second, the HLPE could be caused by the Galactic halo. However, neither can explain the distinct boundaries of the HLPE, which are observed in the second and third Galactic quadrant (cf. thin white lines in Fig.~\ref{sky} top). In fact, synchrotron emission from the halo or local spiral arm should be observable at all latitudes where the local Galactic B-field is perpendicular to the line-of-sight, and not just at the poles. 

\subsection{New Radio Loop}

A previously unknown filament of polarized emission in the southern Galactic hemisphere (hereafter referred to as ``New Loop'') is revealed by the DRAO polarization survey. The New Loop has an excess polarized intensity of $\sim 170$~mK at $1.4$~GHz relative to its surroundings. The elongated shape of the New Loop can be fitted by a small circle. From a fit by eye, the center of this circle is at $l = 345\degr$, $b=0\degr$, with a radius of $65\degr$ (dashed line in Fig.~\ref{sky} top). Faint counterparts can be found in total intensity at $408$~MHz ($\sim 5$~K) and $1420$~MHz ($\sim 250$~mK), which gives a spectral index of $-2.4$. This filament can also be seen in the WMAP polarization data at $23$~GHz.

\subsection{Depolarization Band}
\label{DepolBand}

A minimum of polarized emission from the first Galactic quadrant from $\sim 0\degr$ to  $\sim 70\degr$ at 1.4~GHz, within a band confined by remarkably sharp boundaries at $b \approx \pm30\degr$ (cf. Fig~\ref{sky} top), is seen with great clarity for the first time. The percentage polarization within this area is around $3$\%, compared to $20$ - $30$\% outside the band. Obviously, strong depolarization within the Galactic plane must ``destroy'' the polarized emission from the first Galactic quadrant. The Depolarization Band is prominent in the DRAO data at $1.4$~GHz but not seen in the WMAP data at $23$~GHz, which ephasises the assumption that it is caused by depolarization due to Faraday rotation. The length of the line-of-sight makes differential Faraday rotation a likely depolarization mechanism\footnote{Differential Faraday rotation occurs if synchrotron emitting and Faraday rotating regions are mixed. In this case, polarization vectors with different orientation may superimpose and cancel each other (see also Burn 1966). }. However, the absence of polarized emission from the NPS below $b=30\degr$, in the DRAO data at $1.4$~GHz as well as in the WMAP data at $23$~GHz, cannot be attributed solely to the Depolarization band. It must reflect an inherent property of the emission region. This is explained by the model presented in this paper.

\section{The Model}

\begin{figure}
\includegraphics[width=7.2cm]{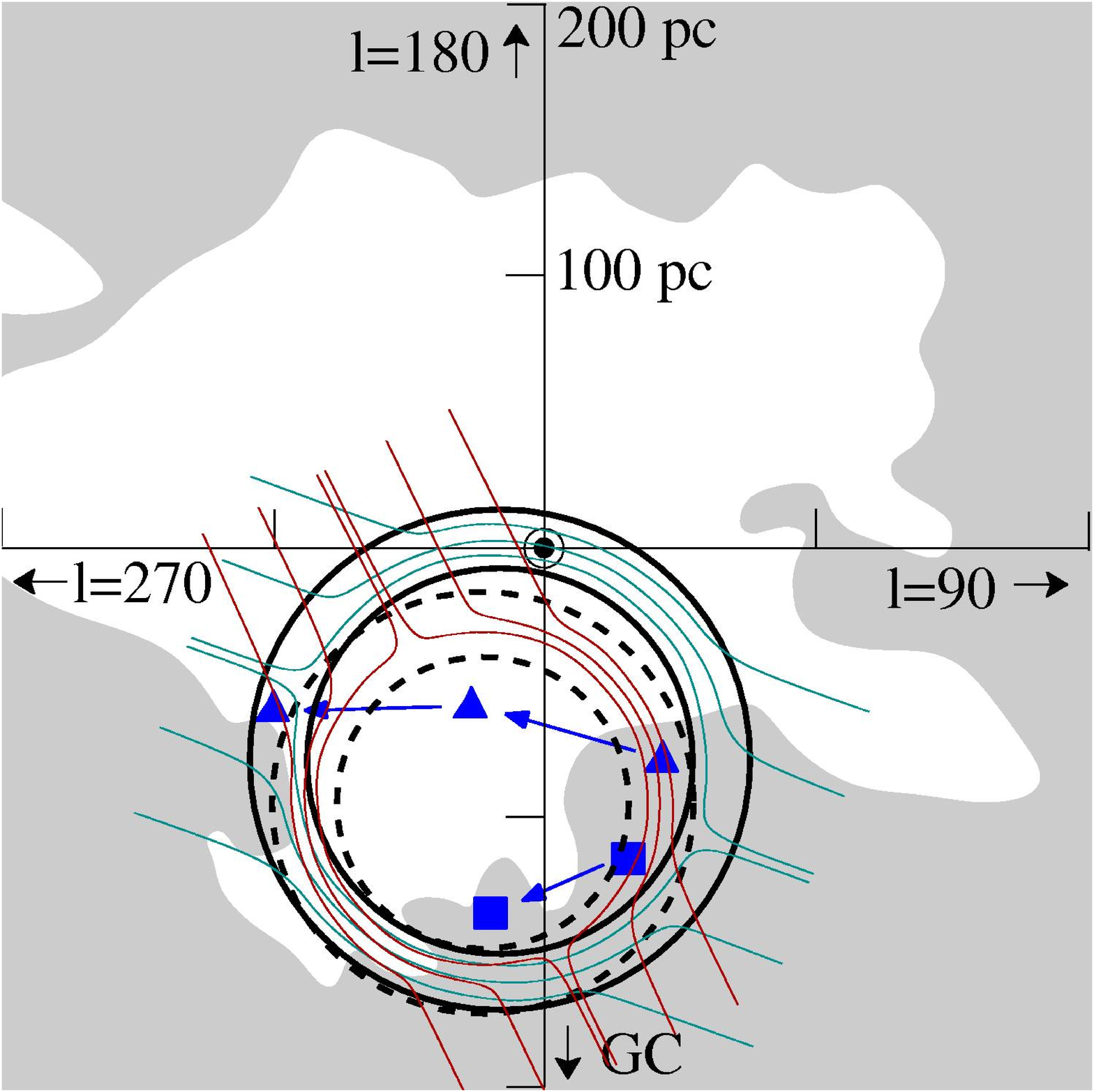} \\
\includegraphics[width=7.2cm]{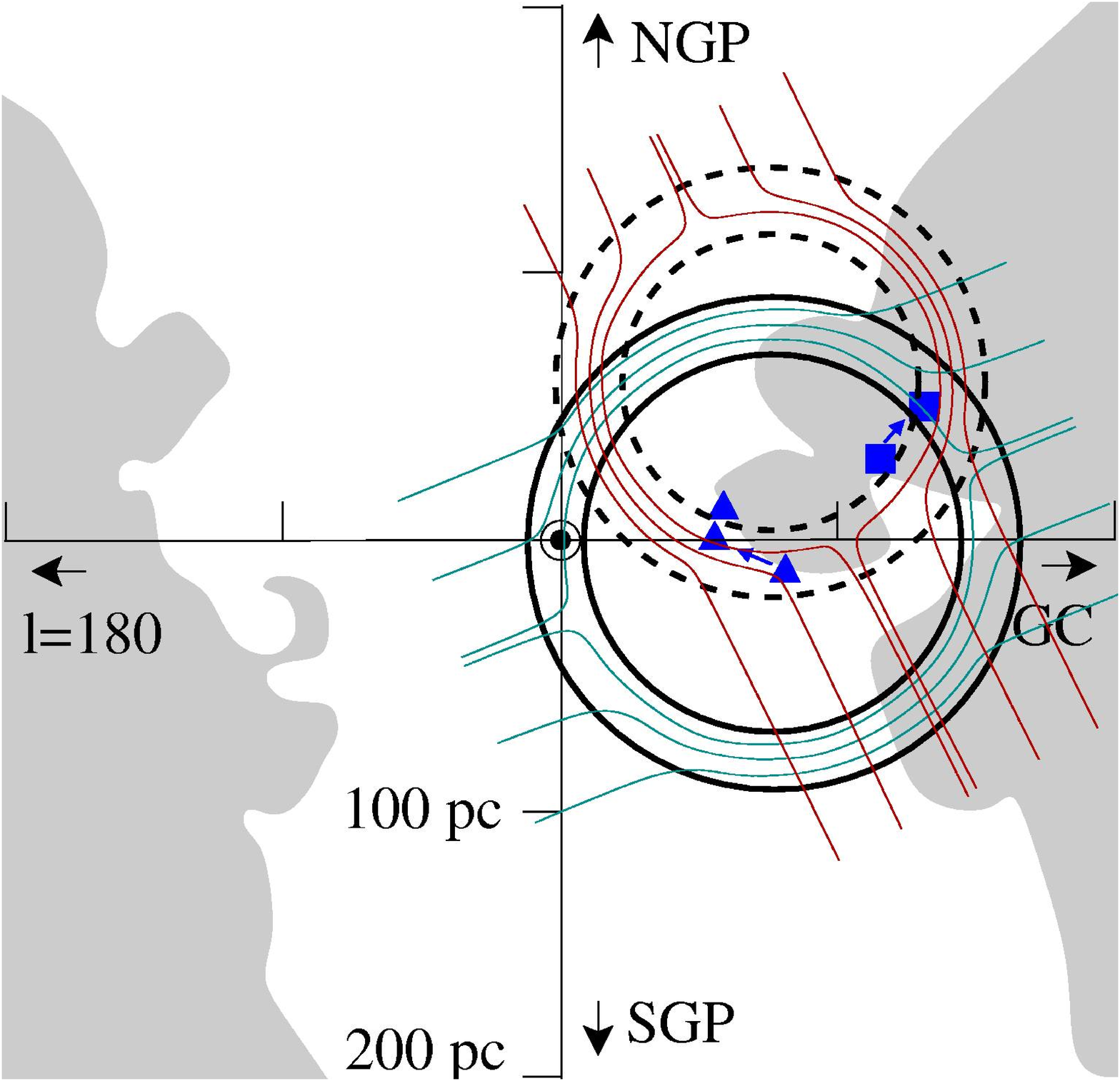}
\caption{The sketch displays cuts through the model viewed looking down on the Galactic plane towards negative latitudes ({\it top}) and the vertical plane through the Sun and perpendicular to the line-of-sight looking towards $l=90\degr$ ({\it bottom}). The two shells are indicated by solid (S1) and dashed (S2) lines. The Sun is indicated by the circled dot. Thin lines indicate the B-field orientation of each shell (the B-field component parallel to the image plane). The stippled region  shows the \ion{Na}{1} distribution around the Local Bubble  \citep[taken from][]{1999A&A...346..785S}. Filled triangles and squares show the centers of the LCC and US subgroup today, 5~Myr ago, and 10~Myr ago (only for LCC) \citep[taken from][]{2001ApJ...560L..83M}.}
\label{3dmod}
\end{figure}

\begin{figure}
\includegraphics[width=8cm]{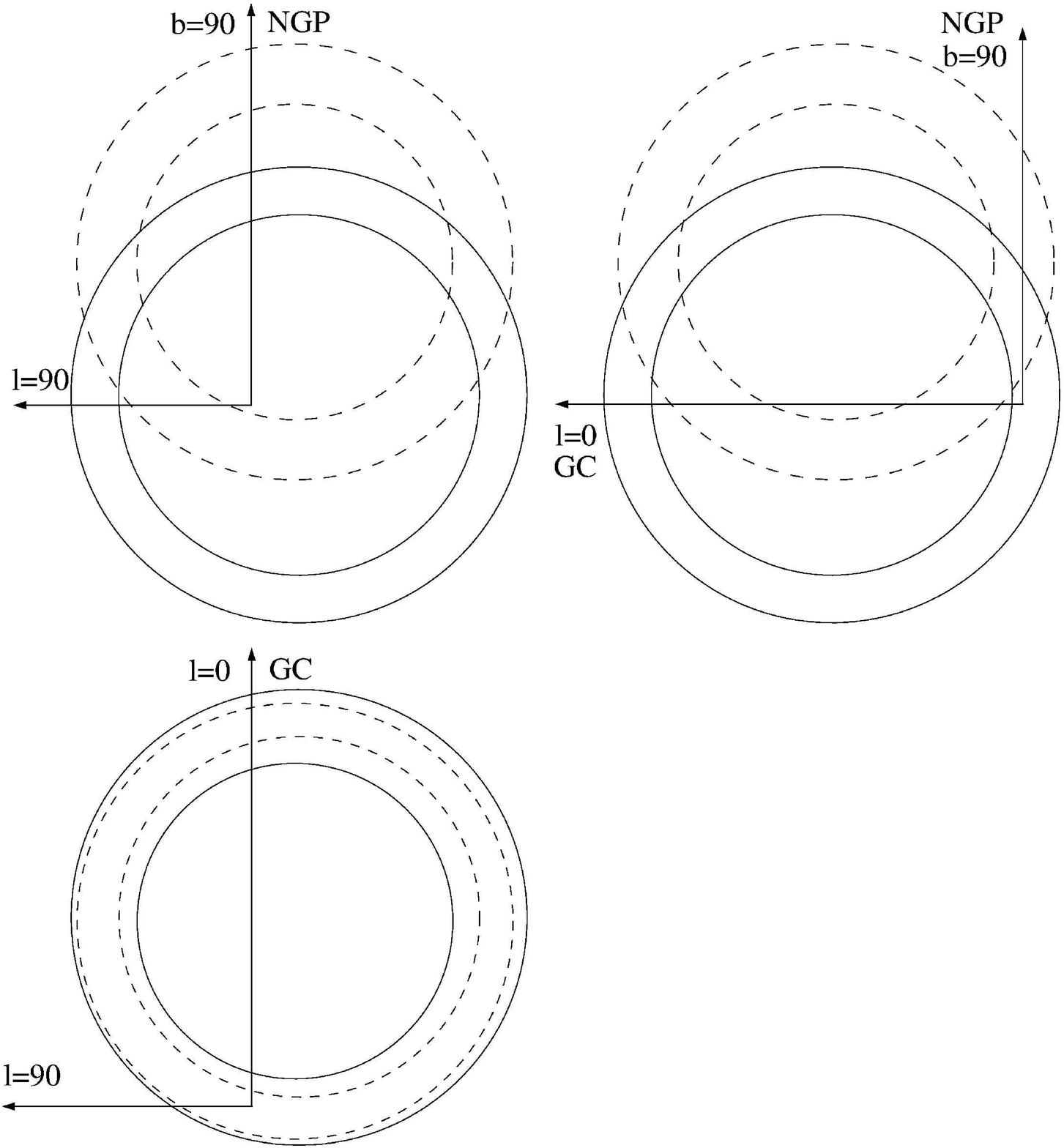}
\caption{An ``engineering drawing'' to help the reader depict the 3-dimensional structure of the two proposed shells. The solid line indicates S1, and the dashed line indicates S2.In contrast to Fig.~\ref{3dmod} these drawings show projections of the two shells rather than cuts through the Galaxy. The Sun is located at the origin of the coordinate system. The figure shows the shells as seen from the Galactic anti-center towards the Galactic center ({\it top left}), seen sideways ({\it top right}), and seen from above ({\it bottom}).}
\label{eng}
\end{figure}

The model consists of two synchrotron emitting shells: S1 and S2. The shells are spherical with constant shell thickness. No emission is produced outside the shells. The Sun resides within S1 between its inner and outer surface (see Fig.~\ref{3dmod} and~\ref{eng}). Theoretical models of magnetic fields in supershells \citep[e.g.][]{1991ApJ...375..239F, 1992PASJ...44..177T} predict that the ambient B-field is pushed by the shock wave and compressed within the expanding shell. The model described here resembles these theoretical B-fields in the simplest way: Magnetic field lines run from the polar caps (the magnetic poles) along longitudes of the shell. The appearance of the  B-field of the shell, that is, its $B_\perp$ and $B_\parallel$ components, depends on the vantage point. In case of a large nearby shell the projection can result in a complicated B-field pattern on the sky.

Each shell is described by 8 parameters (Tab. 1): the center coordinates $l$, $b$, $d$; the inner ($r_{\mathrm{in}}$) and outer ($r_{\mathrm{out}}$) radius of the shell; the angle between the B-field and the line-of-sight  to the  Galactic center ($B_{\theta}$) and to Galactic north ($B_{\phi}$); and two scaling factors describing the intrinsic brightness of each shell. For simplicity, it is assumed that there is no Faraday rotation within the shells. The model described here is an attempt to reproduce the {\it polarized} emission -- the emission tracing the regular B-field -- because this model does not take irregular or turbulent B-field components into account. Therefore, the model must be fitted to polarization maps, as is done here. Moreover, polarization maps at frequencies around $1$~GHz or less are believed to show rather local emission
 
due to depolarization, which helps avoiding confusion with unrelated background emission.

The model uses different field orientations in the two shells. Where the shells overlap in space this is obviously not correct. But the complex shape of the edge of the Local Bubble (cf. Fig~\ref{3dmod}) suggests a very complex evolution, probably the outcome of many earlier stellar winds and supernova events. Any previous large-scale magnetic field has probably been tangled by this complex evolution. Presumably the Local Bubble is not a unique circumstance, so the implication is that most regions in the Galaxy are not characterized by a large-scale quasi-uniform component. The two-field model proposed is a simple approximation to a complex reality, and is unlikely to be correct over the full extent of the shells. However, the good fit to the data demonstrates that it is an adequate model for most of the Loop~I region.

The model was used to calculate the integrated Stokes~$U$ and $Q$ values for each line-of-sight  through the shell complex. Polarized intensity and polarization angle were derived from these integrated values, thereby accounting for depolarization due to superposition of differently oriented polarization vectors along the line-of-sight (see Appendix for a more detailed description). In the northern hemisphere, DRAO data were preferred over WMAP data because of the better sensitivity. Some regions in the DRAO maps were masked out however: the ``Fan-Region'' (see Fig.~\ref{sky}), the two \ion{H}{2}-regions S$\,$27 ($l=4\degr$, $b=22\degr$) and S$\,$7 ($l=350\degr$, $b=22\degr$), and a $\pm 3\degr$ strip along the Galactic plane. In the southern hemisphere, $23$~GHz WMAP data (scaled to $1.4$~GHz using a spectral index of $-3$) were preferred over the 1.4~GHz polarization survey of Testori et al. (2004) because the western caps lie at $b\approx 0\degr$ where depolarization at $1.4$~GHz is likely to be strong. In the WMAP data the Galactic plane was masked out because a simple ``scaling up'' of WMAP data from $23$~GHz to $1.4$~GHz, without taking depolarization effects into account, would result in wrong polarized intensities in regions where Faraday rotation along the line-of-sight is high. The southern Galactic pole in the WMAP was masked out because here sensitivity is too low to permit accurate fitting.

Fitting was done by computer, applying an algorithm that searched for the minimum of the square-root of the sum of the differences between modelled and observed polarized intensities. The algorithm randomly modified model parameters until a good fit was achieved. Different initial start values for the fit were chosen to evaluate the uniqueness of the best-fit. In order to determine the confidence ranges of the best-fit parameters, models with slightly different sets of parameters were tried and visually inspected.

\section{Discussion}

\begin{figure}
\includegraphics[width=15.5cm]{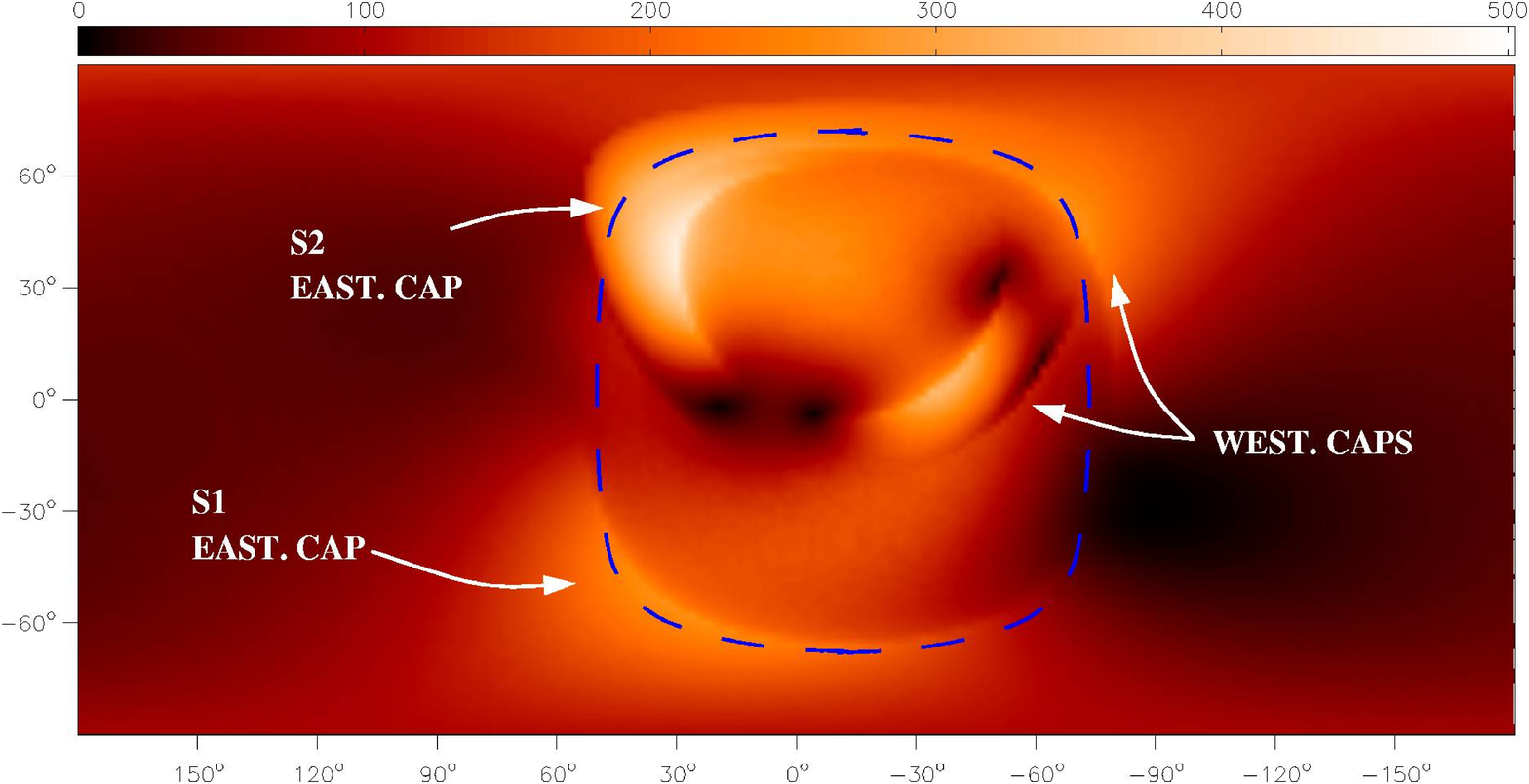}\\
\includegraphics[width=15.5cm]{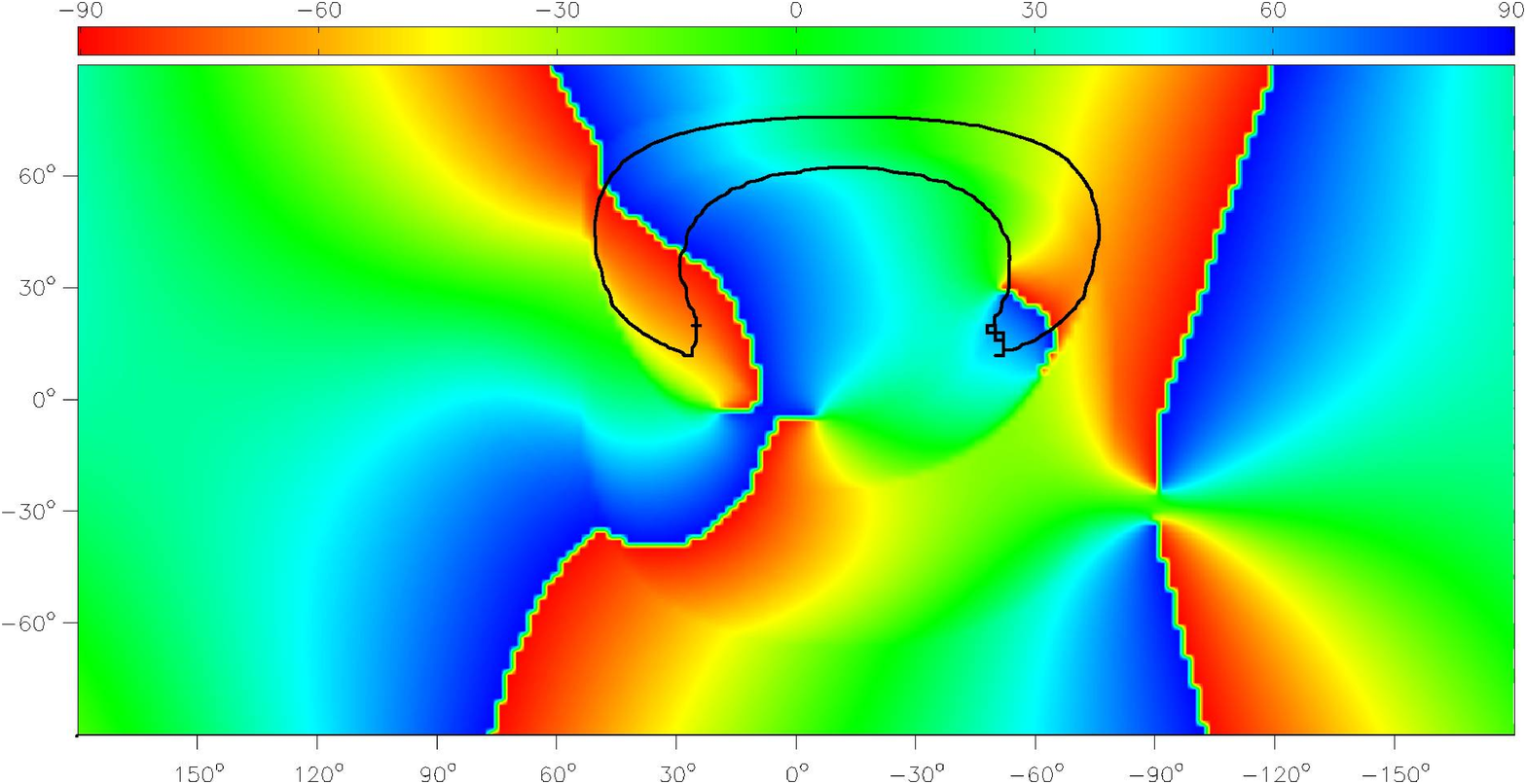}
\caption{Modelled maps of polarized intensity in units of mK ({\it top}) and polarization angle ({\it bottom}). The dashed line from Fig.~\ref{sky} is repeated in the top panel. The black contour in the bottom panel indicates the region where the S1 and S2 shells overlap in space. The model was merely fitted to polarized intensity which means that ``Fig.~\ref{mod} top'' (model) was fitted to ``Fig.~\ref{sky} top'' (survey). Nevertheless, the predicted polarization angles (``Fig.~\ref{mod} bottom'') resemble the observed pattern (``Fig.~\ref{sky} bottom'') remarkably well. The Fan-Region is not part of the model described in this paper. }
\label{mod}
\end{figure}

Figures~\ref{stereo} and \ref{mod} show polarized intensity and polarization angle maps of the best-fit model. At high Galactic latitudes the HLPE is correctly reproduced by S1. Only S1 produces HLPE because the Sun is located {\it inside} this shell, leading to local emission around the Sun. S2 produces the polarized emission of the NPS. At intermediate latitudes, where the line-of-sight through the $B_\perp$ component of S1 is longest, the New Loop is seen. At low latitudes ($|b|\lesssim 30\degr$) synchrotron emission from the two shells is reduced because the path lengths through the shells are short and $B_{\perp}$ is small. Although the model was only fitted to polarized intensities, the polarization angles at intermediate and high Galactic latitudes are remarkably well reproduced. 

The predicted western cap region does not fit the observations as well as the eastern cap region. However, the western caps are likely to be inside the Local Bubble, while the eastern caps may have expanded into a denser medium. Hence, the ``real'' shells are not likely to be of perfectly spherical shape. However, a comprehensive model which includes the full complexity of the ISM around Loop~I is beyond the scope if this paper, and is probably impossible for want of adequate data.

Portions of the two proposed synchrotron shells currently overlap in space, which, projected onto the sky, results in a ring-like region in the northern Galactic hemisphere (see Fig.~\ref{mod} bottom). This region roughly agrees with the location of $1.5$~keV X-ray emission from the NPS.  It is therefore suggested that the X-ray emission is produced by recent interaction of the two shells within this region.

\section{Possible Formation History}

The SCO-CEN association can be divided into three subgroups: Lower Centaurus Crux (LCC), Upper Centaurus Lupus (UCL), and Upper Scorpius (US). \citet{2001ApJ...560L..83M} calculated the positions of the center of each subgroup in the past, taking the effects of solar motion, Galactic rotation, and motions in the $z$-direction into account. Accordingly, the stars whose paths come closest to the center of S1 are those of the LCC\footnote{Note that the scattering of members of these subgroups across the sky means that the UCL is almost as good a candidate.}, which crossed this point $6\pm 2$~Myr ago at a distance of $70$~pc from the Sun. Closest to the center of S2 is the US group at its current location ($l=350\degr$, $b=20\degr$, $145$~pc away from the Sun). Stellar activity within the LCC subgroup started $11$ to $12$~Myr ago, and $5$ to $6$~Myr ago within the US \citep{1989A&A...216...44D}.

The picture that emerges from these data can be summarized as follows. The first supernovae are expected to take place about $3$-$5$~Myr after formation of an association. Thus, about $7$ to $8$~Myr ago, supernova explosions began to occur within the LCC and, subsequently, delivered enough energy to inflate a bubble around LCC -- the S1-shell. The US, instead, has only recently reached its evolutionary time scale and has just begun inflating the Loop~I bubble (S2). The shock front of Loop~I (S2) hit the LCC bubble (S1) just recently ($10^4$ years ago or less), giving rise to the X-ray emission observed toward the NPS. 

In this picture S1 is about $6$~Myr old, based on the positional coincidence with the LCC at this time. S2 is $1$-$2$~Myr old, based on the the age of the US. This suggests that S1 is more evolved than S2, which may explain why almost no observational tracers of S1 can be found in total intensity surveys. S1 seems to be an almost dissolved shell, whose last observable remnants are the HLPE and the New Radio Loop.

The scenario proposed here has compelling similarities to the models of \ion{H}{1} shells in this region by \citet{1979IAUS...84..295W} and \citet{1987Ap&SS.138..229B}, mentioned earlier in the Introduction. The ages derived above and best-fit radii of the model give expansion velocities of $8$ and $32$~km$\,$s$^{-1}$ for S1 and S2, respectively, using the standard model for the kinematic age of stellar wind bubbles with $t_{\mathrm{kin}}=0.6\,R/v_{\mathrm{exp}}$ \citep{1977ApJ...218..377W}. Taking into account the uncertainties in this estimate, these velocities agree with the velocities of $2$~km$\,$s$^{-1}$ \citep{1979IAUS...84..295W} and $19$-$25$~km$\,$s$^{-1}$ \citep{1984ApJS...55..585H, 1974PASJ...26..399S} found for \ion{H}{1} gas in this region. This may imply that the \ion{H}{1} distribution in this region is the result of two superimposed \ion{H}{1} shells that are expanding with different velocities.

\section{Conclusions}

Based on an analysis of the new DRAO polarization survey and scaled WMAP data, a model consisting of two synchrotron emitting shells is developed that reproduces large-scale structures in the polarized sky. One of these shells, S1, has reached the Sun and gives rise to polarized synchrotron emission at high Galactic latitudes, the HLPE described in this paper. Where the path length through S1 is longest and its B-field is perpendicular to the line-of-sight over its whole length in the emitting region, a new radio loop is detected in polarized intensity. A younger shell, S2, correctly reproduces emission from the NPS. A scenario is proposed in which S2 has recently started to interact with S1, possibly giving rise to the observed X-ray emission. The model also predicts correctly the low polarization from S1 and S2 within the Depolarization Band. The picture of the local ISM suggested in this paper is in agreement with previous studies outlined in the Introduction, although the two-shell geometry of the model makes a comparison with previous models difficult.

\acknowledgments
The author would like to thank T. L. Landecker, E. M. Berkhuijsen, R. Kothes, A. D. Gray, and P. Vaudrevange for comments on the manuscript. The Dominion Radio Astrophysical Observatory is a National Facility operated by the National Research Council Canada. The Canadian Galactic Plane Survey is a Canadian project with international partners, and is supported by the Natural Sciences and Engineering Research Council (NSERC). I acknowledge the use of the Legacy Archive for Microwave Background Data Analysis (LAMBDA). Support for LAMBDA is provided by the NASA Office of Space Science.

\appendix
\section{Magnetic Field Used In The Model}

\begin{figure}
\begin{center}
\includegraphics[width=8.5cm]{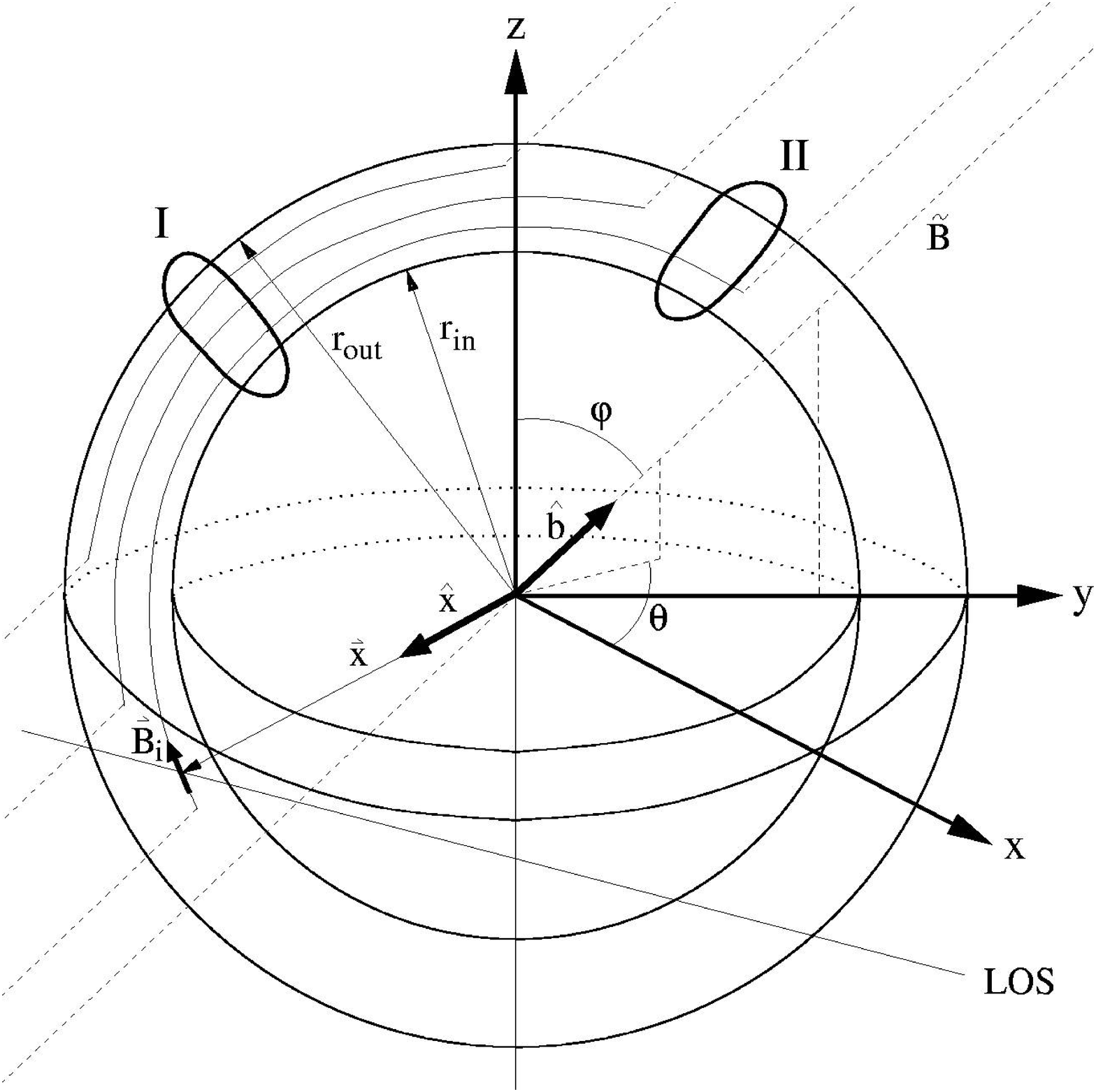}
\caption{Geometry of the spherical shell. $r_{\mathrm{in}}$ and
  $r_{\mathrm{out}}$ are the inner and outer radii of the shell, respectively;
  $\tilde{B}$ is the background magnetic field and $\hat{b}$ is its unit
  vector; $\vec{B}_i$ is the magnetic field inside the shell along the
  i-th line-of-sight element; and $\hat{x}$ is the unit vector pointing
  towards the $i$-th line-of-sight element. Region I and II indicate regions with strong and weak B-fields,
   respectively.}
\label{sketch}
\end{center}
\end{figure}

In this model an approximation for the magnetic field $\vec{B_i}$
inside an expanding shell is used. Its strength and direction depend on the direction ($\theta$, $\varphi$) of the ambient magnetic field $\tilde{B}$ as well as the inner and outer radii of the shell $(r_{\mathrm{in}}, r_{\mathrm{out}})$. Looking at Fig.~\ref{sketch}, it is easy to see that the magnitude of $\vec{B_i}$ is maximal / zero where the expansion is perpendicular / parallel to $\tilde{B}$. For $|\vec{x}| > r_{\mathrm{out}} $ and $|\vec{x}| < r_{\mathrm{in}}$ the magnetic field strength $\vec{B_i}$ is assumed to be zero.

The direction of $\vec{B_i}$ is given by
\begin{equation}
    \hat{B_i} = \frac{\vec{\rho}_i}{|\vec{\rho}_i|},
\end{equation}
where $\vec{\rho}_i = \hat{b}-(\hat{b}\cdot\hat{x}) \hat{x}$ with $\hat{b}$ the unit vector of the ambient field $\tilde{B}$ and $\hat{x}$ the unit vector pointing towards the $i$-th line-of-sight element. The field strength of $\vec{B_i}$ can be evaluated to
\begin{equation}
    n_i = (\hat{b}\cdot \hat{x})^2,
\end{equation}
giving rise to the Stokes parameters 
\begin{equation}
\begin{array}{lll}
u_i &\propto& n_i\,B_{\perp}^{(\gamma+1)/2}\,\sin\left[2\,\arctan\left(\frac{B_l}{B_b}\right)\right],\\
q_i &\propto& n_i\,B_{\perp}^{(\gamma+1)/2}\,\cos\left[2\,\arctan\left(\frac{B_l}{B_b}\right)\right], 
\end{array}
\end{equation}
for the $i$-th line-of-sight. Here $B_\perp=\sqrt{B_l^2+B_b^2}$ and $B_l, B_b$ are the projections of $\vec{B_i}$ onto the line-of-sight. An energy spectral index of the synchrotron radiation of $\gamma\approx 2.8$ is taken. Assuming no intrinsic Faraday rotation, the polarized intensity and the polarized position angle is finally found
\begin{equation}
\begin{array}{lll}
PI &=& \sqrt{U^2 + Q^2},\\
PA &=& \frac{1}{2}\,\arctan\left(\frac{U}{Q}\right),
\end{array}
\end{equation}
where $U=\sum u_i$ and $Q=\sum q_i$.

\clearpage

\begin{deluxetable}{lccccccc}
\tabletypesize{\scriptsize}
\tablecaption{Best-fit parameters of the model.}
\tablewidth{0pt}
\tablehead{
\colhead{} & \colhead{$l$} & \colhead{$b$} & \colhead{$d$} & \colhead{$r_{\mathrm{in}}$} & \colhead{$r_{\mathrm{out}}$} & \colhead{$B_{\theta}$} & \colhead{$B_{\phi}$}\\
\colhead{Shell} & \colhead{(deg)} & \colhead{(deg)} & \colhead{(pc)} & \colhead{(pc)} & \colhead{(pc)} & \colhead{(deg)} & \colhead{(deg)}}
\startdata
  S1 &  $346$ &  $3$ & $78$ & $72$ & $91$ & $71$ & $-72$\\
    &  $\pm 5$ &  $\pm 5$ & $\pm 10$ & $\pm 10$ & $\pm 10$ & $\pm 30$ & $\pm 30$\\
  \vspace{-1.5mm}\\
  S2 &  $347$ &  $37$ & $95$ & $63$ & $87$ & $25$ & $25$\\
    &  $\pm 15$ &  $\pm 15$ & $\pm 10$ & $\pm 15$ & $\pm 10$ & $\pm 30$ & $\pm 30$
   \enddata
\label{freeparameters}
\end{deluxetable}

\clearpage

\clearpage

\clearpage

\clearpage

\clearpage

\clearpage

\end{document}